\documentclass[useAMS]{mn2e}
\usepackage{rotating}
\long\def\symbolfootnote[#1]#2{\begingroup%
\def\thefootnote{\fnsymbol{footnote}}\footnote[#1]{#2}\endgroup}
\newcommand{\gae}{\lower 2pt \hbox{$\, \buildrel {\scriptstyle >}\over {\scriptstyle
\sim}\,$}}
\newcommand{\lae}{\lower 2pt \hbox{$\, \buildrel {\scriptstyle <}\over {\scriptstyle
\sim}\,$}}
\begin{document}

\title[Scattered Emission from Relativistic Outflow]
{Scattered Emission from A Relativistic Outflow and Its Application to Gamma-Ray Bursts}

\author[Shen, Barniol Duran \& Kumar]{R.-F. Shen$^{1}$\thanks
{E-mail: rfshen@astro.as.utexas.edu; rbarniol@physics.utexas.edu; pk@astro.as.utexas.edu},
R. Barniol Duran$^{1, 2}$\footnotemark[1] and P. Kumar$^{1}$\footnotemark[1]\\
$^{1}$Department of Astronomy, University of Texas at Austin, Austin, TX 78712, USA\\
$^{2}$Department of Physics, University of Texas at Austin, Austin,
TX 78712, USA}

\date{Accepted  2007;
      Received 2007;
      in original form  2007 July 21}

\pagerange{\pageref{000}--\pageref{000}} \pubyear{2007}

\maketitle

\begin{abstract}
We investigate a scenario of photons scattering by electrons within
a relativistic outflow. The outflow is composed of discrete shells
with different speeds. One shell emits radiation for a short
duration. Some of this radiation is scattered by the shell(s)
behind. We calculate in a simple two-shell model the observed
scattered flux density as a function of the observed primary flux
density, the normalized arrival time delay between the two emission
components, the Lorentz factor ratio of the two shells and the
scattering shell's optical depth. Thomson scattering in a cold shell
and inverse Compton scattering in a hot shell are both considered.
The results of our calculations are applied to the Gamma-Ray Bursts
and the afterglows. We find that the scattered flux from a cold
slower shell is small and likely to be detected only for those
bursts with very weak afterglows. A hot scattering shell could give
rise to a scattered emission as bright as the X-ray shallow decay
component detected in many bursts, on a condition that the
isotropically equivalent total energy carried by the hot electrons
is large, $\sim 10^{52-56}$ erg. The scattered emission from a
faster shell could appear as a late short $\gamma$-ray/MeV flash or
become part of the prompt emission depending on the delay of the
ejection of the shell.
\end{abstract}


\section{Introduction}

Gamma-Ray Bursts (GRBs) are a cosmological phenomenon with a huge
energy release, fast variabilities and very complex multi-wavelength
light curves. A relativistic outflow is unavoidable in order to
explain the fast variability and so called ``compactness problem''
(see Piran 2005 for a review). According to the standard
``Fireball'' model, the outflow from the GRB central engine has a
finite duration and can have a wide range in its velocities,
thus can be modeled by being made of discrete relativistic shells.
These shells are responsible for the observed $\gamma$-rays (via
internal shocks) and for the afterglow emissions (via external
shocks) (cf. Piran 2005). In this picture, if one
shell emits $\gamma$-rays, some fraction of that emission should be
scattered by shells behind, and the scattered emission would arrive at
the observer at a different time, with a different flux and possibly
at a different photon frequency.  Detection of the scattered photons
would help us explore the properties of the GRB ejecta and the late
outflow.

In the present paper we consider a simple scenario, where only two
consecutive shells are present: one shell radiates and the other
receives some of this radiation and scatters it. The two shells can
have different speeds and the shell that receives and scatters the
emission can have an arbitrarily large time delay in its ejection
from the central source, but it has to be behind the emitting shell.
An observer detects the primary emission from the
first shell and then the scattered emission from the second one at a
later time because of the light-travel time. We calculate the ratio
between these two emissions' fluxes, the time delay in their
arrival, the ratio between their frequencies and the ratio between
their durations.

Early GRB X-ray observations by {\it Swift} have shown a
``canonical'' behavior that presents a puzzling shallow decay typically
lasting for a few hours (e.g. Nousek et al. 2006).
This decay phase is poorly understood (see Zhang 2007 for a review of
current possible models).  We will explore the possibility that this
shallower decay could be due to the scattered emission.

The scattering of the GRB prompt emission photons by electrons or
dust grains in a dense circum-burst environment has been
investigated before (e.g., Esin \& Blandford 2000; Madau et al.
2000; Shao \& Dai 2007; Heng et al. 2007). The scattering process we
consider in this work happens within the GRB outflows, which is a
natural outcome of the outflow when it has a finite duration and a
variable speed.

The paper is structured as follows. We first derive a general
formula for the observed flux from a relativistic shell in \S2. In
\S3 we construct the flux and geometrical relations for a two-shell
model. The formulae for the scattering process are developed in \S4.
Then we elaborate the primary and scattered emission relations such
as time delay, frequency ratio and time duration ratio
in \S5. We present the main result - the ratio between the scattered
and the primary fluxes - in \S6.  The application to the GRB shallower
decay data is presented in \S7. A faster scattering shell case
is discussed in \S8.  We also discuss X-ray dim bursts and X-ray-dark
short bursts, for which the scattered emission might be easier to
detect, in \S9.  Finally, the summary and conclusions are given in \S10.

\section{Emission from a relativistic shell}

Consider a spherical shell moving relativistically with Lorentz
factor (LF) $\Gamma$ (when the shell is beamed with an opening angle $\ge
\Gamma^{-1}$, it still can be considered as being spherical). The
surface brightness in the rest frame of the shell is
$\epsilon'_{\nu'}$ (erg s$^{-1}$ cm$^{-2}$ Hz$^{-1}$ sr$^{-1}$), the
luminosity distance between the observer and the shell is $D_L$ and the
radius of the shell is $R$, both of these distances measured in the
laboratory frame.  The flux density received at a frequency $\nu$ by the
observer ahead of the shell, $f_\nu$, can be obtained by calculating
the specific luminosity of the relativistic shell. The luminosity of
the shell is given by $f_\nu (4 \pi D_L^2) = \epsilon'_{\nu'}(4 \pi R^2)
\Gamma (2 \pi)$.  The last expression
includes a factor of $\Gamma$, to take into account the ``boost''
that the photons experience; and a factor of $2 \pi$, since we
assume that the photons are being emitted isotropically from the
shell, in the rest frame of the shell, and we are only interested
in the ones reaching the observer.  The two expressions yield

\begin{equation}
f_\nu = 2 \pi \epsilon'_{\nu'} \Gamma \biggl(\frac{R}{D_L}\biggr)^2.
\end{equation}

\section{Two shells scenario}

Consider now two thin shells being ejected with an half opening
angle of $\theta_j$ from the central engine. Shell 1 is ejected
first with LF $\Gamma_1$ and, after a delay $\delta t$, measured in
the laboratory frame, shell 2 is ejected with LF $\Gamma_2$. We
assume that shell 1 is emitting photons isotropically in its
co-moving frame and is characterized by an angular-independent
surface brightness, $\epsilon'_{\nu'}$, on both sides of the shell.
In the laboratory frame most of these photons will appear to move in
the same direction as shell 1 and reach a distant observer, and a
few will move in the opposite direction, encountering shell 2 on
their way. These photons will be scattered by shell 2 and then reach
the observer. See Figure 1 for an illustration. We will use primed
quantities to specify the co-moving frame where the quantity is
being measured: unprimed corresponds to the laboratory frame,
primed ($'$) to the co-moving frame of shell 1, and double primed
($''$) to the co-moving frame of shell 2.

Before proceeding with the detailed calculations about the scattered
emission, we provide simple scaling relationships between the observed
scattered emission and the observed direct emission from shell 1 by only
considering the line-of-sight region.

Let us assume $\nu$ is the photon frequency of the direct emission from
shell 1. In the shell 1 co-moving frame, the emitted photon frequency is
$\nu' \simeq \frac{\nu}{\Gamma_1}$ due to the relativistic Doppler effect
(for simplicity we neglect the factor of 2). As seen by shell 2 the photon
has a frequency of $\nu'' \simeq \nu'\frac{\Gamma_2}{\Gamma_1}$. If shell 2
is cold, the scattering does not change the photon's energy. The observed
frequency of the scattered emission is $\nu_s \simeq \nu''\Gamma_2$. Thus
the observed frequency ratio between the two emissions is
$\frac{\nu_s}{\nu} \simeq (\frac{\Gamma_2}{\Gamma_1})^2$.

The scattered emission will be observed at a later time, because the scattered
photon has traveled an extra distance. This extra distance is equal to
$R_1(1-\frac{\beta_2}{\beta_1})+\beta_2 \delta t$, where $R_1$ is the distance
of shell 1 from the central engine when a photon was emitted from shell 1
toward shell 2, and $\beta_1$ and $\beta_2$ are shell 1 and shell 2 velocities,
respectively (the first part of the expression is due to the difference in the
speeds of the shells, and the second part is due to the ejection delay of
shell 2). The observed delay of the scattered emission is this separation
divided by the speed of light
$\approx \frac{R_1}{2\Gamma_2^2 c} \approx T(\frac{\Gamma_1}{\Gamma_2})^2$
for small $\delta t$, where $T$ is the observed time of the shell 1
direct emission since the central engine explosion.

If $L$ is the luminosity observed directly from shell 1, the shell 1 co-moving
frame luminosity would be $L' \simeq \frac{L}{\Gamma_1^2}$. In the
shell 2 co-moving frame, shell 1 has a luminosity of $L''$.
Using the Lorentz invariance of $I_{\nu}/\nu^3$ we find that the luminosity
ratio, $\frac{L''}{L'}$, is the frequency ratio to the fourth power, or
$\frac{L''}{L'} \simeq (\frac{\nu''}{\nu'})^4 \simeq (\frac{\Gamma_2}{\Gamma_1})^4$.
In the shell 2 co-moving frame,
$\tau_e L''$ is the luminosity of the scattered emission, where
$\tau_e$ ($<1$) is the shell 2 electron's optical depth. Then the observed
luminosity of the scattered emission is $L_s \simeq \tau_e L'' \Gamma_2^2$.
Thus we obtain that $\frac{L_s}{L} \simeq \tau_e (\frac{\Gamma_2}{\Gamma_1})^6$,
which shows that the observed luminosity ratio is strongly dependent on the
LF ratio. Since $\frac{\nu_s}{\nu} \simeq (\frac{\Gamma_2}{\Gamma_1})^2$,
this indicates that the observed specific flux ratio is
$\frac{f_{\nu,s}^s}{f_{\nu}} \simeq \tau_e (\frac{\Gamma_2}{\Gamma_1})^4$.

\begin{figure*}
\centerline{\hbox{\includegraphics[width=9cm,angle=0]{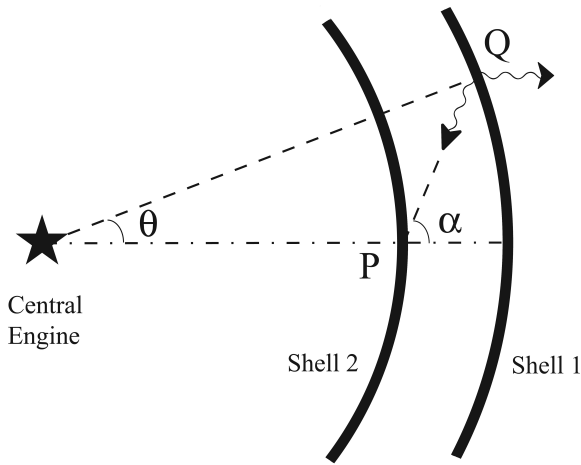}}}
\caption {A simple two-shell scenario geometry. The diagram shows
two photons emitted at the same time from point $Q$ on shell 1. One
photon travels to the observer ahead of shell 1, while the other
travels back to point $P$ on shell 2. Note that the figure shows
shell 1 and the photons at time $t$ and shell 2 at time $t+l$, where
$l$ is the light travel time from $Q$ to $P$.}
\end{figure*}

\subsection{Incident flux on shell 2}

Now we work on the calculation of the scattered flux on more detail.
In order to determine the scattered flux from shell 2 we will first
calculate the incident flux from shell 1 at the point of shell 2
that intersects the line of sight between the central engine and the
observer - we will call this point $P$.  To do this, we will use the
Lorentz invariance of $I_{\nu}/\nu ^3$, where $I_{\nu}$ (erg
s$^{-1}$ cm$^{-2}$ Hz$^{-1}$ sr$^{-1}$) is the specific intensity
and $\nu$ is the frequency of the photon.

Let us consider a bundle of rays being emitted at an arbitrary
point - $Q$ - on shell 1 and directed to the point $P$ (see Figure
1). The angle between the line of sight and the line connecting the
central engine and $Q$ is $\theta$. The angle between the line of
sight and the line connecting $P$ and $Q$ is $\alpha$. From the
Lorentz invariance of $I_{\nu}/\nu ^3$ we can obtain the relation

\begin{equation}
\frac{I_{\nu}}{I''_{\nu''}}= \biggl[\frac{1}{\Gamma_2(1 + \beta_2 \cos \alpha ) }   \biggr]^3,
\end{equation} where $I_{\nu}$ is the specific intensity of the
bundle of rays measured by an observer standing still in the
laboratory frame and at the position of point $P$ (note that
$I_{\nu}$ is NOT the specific intensity of the emission detected by
a distant laboratory-frame observer sitting in front of shell 1);
$I''_{\nu''}$ is the specific intensity of the bundle of rays
measured in the shell 2 co-moving frame at point $P$. We can
relate $I_{\nu}$ with the specific intensity of the bundle of
rays measured in the shell 1 co-moving frame at point $Q$,
$I'_{\nu'}$, as follows

\begin{equation}
\frac{I_{\nu}}{I'_{\nu'}}= \biggl\{\frac{1}{\Gamma_1 [1 + \beta_1 \cos( \alpha - \theta)] }   \biggr\}^3.
\end{equation}

We need to obtain a relation between the co-moving specific
intensity, $I'_{\nu'}$, and the surface brightness,
$\epsilon'_{\nu'}$, of shell 1, both quantities in shell 1 co-moving
frame.  This relation is given by
\begin{equation}
I'_{\nu'} = \frac{ \epsilon'_{\nu'} }{ \cos \eta' }
\end{equation}
where $\eta'$ is the angle measured in the shell 1 co-moving frame
between the photon's direction and the normal to the emitting
surface (facing shell 2). $\eta'$ can be determined using the {\it
aberration of light} formula (Rybicki \& Lightman 1979):

\begin{equation}
\cos \eta' = \frac{ \cos(\alpha - \theta) + \beta_1 }{1 + \beta_1 \cos(\alpha - \theta) }
\end{equation} where all the quantities have been defined previously.

Finally, using formulas (2) - (5) one can obtain the incident flux from
shell 1 on point $P$ in shell 2 co-moving frame, $f''_{\nu''}$, which
is given by
\begin{displaymath}
 f''_{\nu''} = \int{ I''_{\nu''} \cos \alpha''\,d\Omega''}
= 2 \pi \int{ I''_{\nu''} \cos \alpha'' \sin \alpha'' \,d\alpha''},
\end{displaymath}
or
\begin{equation}
 f''_{\nu''} = \pi \epsilon'_{\nu'} \biggl(\frac{\Gamma_2}{\Gamma_1} \biggr)^3
\int{\frac{(1+\beta_2\cos\alpha)^3\,d\sin^2\alpha''}
{[\beta_1+\cos(\alpha-\theta)] [1+\beta_1\cos(\alpha-\theta)]^2}},
\end{equation}
where the integral runs from $0$ to {\bf$ \alpha_j'' $} -  the half
opening angle of shell 1 as seen from an observer on point $P$ in
shell 2 (the subscript ``j'' denotes the edge of shell 1).

\subsection{Light path geometry}

In order to solve this last integral, we need to understand the
relation between the angles and how they transform in different
frames.  First, we need to setup the geometry of the
problem. The photon that is emitted from point
$Q$ at time $t$ travels a distance $l$ and reaches point $P$ on
shell 2 at time $t+l$ (here, and throughout the paper, we use units
in which the speed of light is $1$). The radius at which the photon
was emitted from point $Q$ on shell 1 is $R_1(t)$ and reaches point
$P$ on shell 2 at radius $R_2(t+l)$, where $R(t)$ can be obtained by
$R(t)=\beta t$ (all these quantities are measured in the laboratory
frame). It is important to remember that there is a time delay in
the ejection of shell 2, which will be taken into account when using
$R_2(t)$. The geometry describing the light path of the photon gives:
\begin{equation}
l \cos \alpha = R_1(t)-R_2(t+l)
\end{equation}
\begin{equation}
l \sin \alpha = R_1(t) \sin\theta,
\end{equation}
from which it can be shown that
\begin{equation}
\tan\alpha = \frac{\theta}{1-\frac{\beta_2}{\beta_1}+
  \frac{\beta_2 \delta t}{R_1}-\frac{\beta_2\theta}{\sin\alpha}},
\end{equation}
where we use $R_1$ to denote $R_1(t)$ and we have also used the fact
that $\theta$ is small, so that $\sin\theta\simeq\theta$. And again,
using the {\it aberration of light} formula we have
\begin{equation}
\tan \alpha'' = \frac{\sin \alpha}{\Gamma_2(\cos \alpha + \beta_2)}.
\end{equation}

Also from (9) and (10) we can get a close form expression for $\alpha''$:
\begin{equation}
\tan\alpha''= \frac{\theta}{\Gamma_2(1-\frac{\beta_2}{\beta_1}+\frac{\beta_2\delta t}{R_1})}.
\end{equation}

Equations (10) and (11) allow us to carry out the integral in equation (6).
Careful analysis (see Appendix) shows that the integrand is weakly dependent
on $\alpha''$ and it reduces to a constant of order unity.
This simplifies $f''_{\nu''}$ greatly and we obtain:
\begin{equation}
f''_{\nu''} = \pi \epsilon'_{\nu'} \biggl(\frac{\Gamma_2}{\Gamma_1} \biggr)^3
\int{\,d\sin^2\alpha''} = \pi \epsilon'_{\nu'} \biggl(\frac{\Gamma_2}{\Gamma_1} \biggr)^3 \sin^2\alpha_j''
\end{equation}
where $\alpha_j''$ can be obtained from equation (11) by setting $\theta=\theta_j$.

\section{Scattering from shell 2}

\subsection{Scattered flux}

Assuming that shell 2 is ``cold'' and knowing the flux from shell 1 at
point $P$, we can calculate the scattered flux from shell 2.  We will
assume that the photons from shell 1 undergo {\it Thomson scattering}
with the electrons on shell 2 and that the scattering
is isotropic in the co-moving frame.  The surface brightness of shell 2 in
its co-moving frame will be given by
\begin{displaymath}
\epsilon''_{\nu''} = \frac{1}{4\pi} \Sigma_e \sigma_T  f''_{\nu''},
\end{displaymath}
where $\Sigma_e$ is
the electron surface density and $\sigma_T$ is the Thomson
scattering cross section. To obtain the scattered flux at the
scattered frequency that reaches an observer at a luminosity
distance $D_L$ one can use equation (1)
\begin{displaymath}
f^{s}_{\nu,s} = 2 \pi \epsilon''_{\nu''} \Gamma_2 \biggl[\frac{R_2(t_{scat})}{D_L}\biggr]^2
\end{displaymath}
and obtain
\begin{equation}
f^{s}_{\nu,s} = \frac{1}{2} \tau_e \Gamma_2  \biggl[
\frac{R_2(t_{scat})}{D_L} \biggr]^2 f''_{\nu''}
\end{equation}
where we have $\tau_e=\Sigma_e \sigma_T$, the optical depth for electrons
to Thompson scattering. We also have used $R_2(t_{scat})$ to specify that
the radius of the second shell needs to be calculated at a later time,
$t_{scat}$, when the scattering occurs\footnote{We approximate the
incident flux on point P to the incident
flux on any other point on shell 2. The validity of this
approximation depends on the magnitude of $\alpha_j''$.
If $\alpha_j''$ is small, the angular size of shell 2 as seen
by an observer co-moving with shell 2 at the point where the
line of sight intersects with shell 1 - let us call it $\Lambda_j''$
- is also small, because
$\frac{\Lambda_j''}{\alpha_j''}=\frac{R_2(t_{scat})}{R_1(t)} < 1$;
then this approximation should be good. For our interested parameter
space - determined from the GRB data we are going to use -
$\alpha_j''$ is between 0.1 and $\frac{\pi}{4}$ (see \S 7.1.2), not
very small. Thus the approximation overestimates the scattered flux
slightly.}.  Using equation (12) we obtain
\begin{displaymath}
f^{s}_{\nu,s} = \frac{\pi}{2} \tau_e \epsilon'_{\nu'}
\biggl(\frac{\Gamma_2^4}{\Gamma_1^3} \biggr)
\biggl[\frac{R_2(t_{scat})}{D_L}\biggr]^2  \sin^2\alpha_j'',
\end{displaymath}
which allows us to find a ratio between the scattered flux from shell 2,
$f^{s}_{\nu,s}$, and the direct flux from shell 1, $f_{\nu}$,

\begin{equation}
\frac{f^{s}_{\nu,s}}{f_{\nu}}= \frac{1}{4} \tau_e \sin^2 \alpha''_j
\biggl[\frac{R_2(t_{scat})}{R_1(t)}\biggr]^2 \biggl(
\frac{\Gamma_2}{\Gamma_1}\biggr)^4,
\end{equation}
where we have used equation (1) to immediately get $f_{\nu}$.
It is important to note that these two fluxes arrive at different
times. This will be further explained in detail in the next section.
Also, the result for $\tau_e > 1$ is the same as that for
$\tau_e=1$.

Recall that before the detailed calculation we estimated the
observed specific flux ratio under the line-of-sight approximation.
Eq. (14) is consistent to what we estimated earlier, except that
the previously ignored shell solid angle term is fully considered here.

\subsection{Shell radii}

All quantities of the flux ratio are known, except the ratio of the
two radii. To obtain it, we will consider a photon emitted on shell
1 that travels along the line of sight to shell 2.   The light
travel time of this photon is given by equation (7) but using $\alpha=0$,
so that
\begin{displaymath}
l=R_1(t)-R_2(t+l).
\end{displaymath}
Solving for $l$ we obtain

\begin{equation}
l=\frac{R_1(t)-R_2(t)}{1+\beta_2}.
\end{equation} The ratio of the radii is given by

\begin{displaymath}
\frac{R_2(t_{scat})}{R_1(t)}=\frac{\beta_2(t-\delta t + l)}{\beta_1 t},
\end{displaymath}
and using equation (15) and the fact
that both shells move close to the speed of light, we find

\begin{equation}
\frac{R_2(t_{scat})}{R_1(t)}=\frac{\beta_2}{\beta_1}-\frac{\beta_2\delta t}{2 R_1}.
\end{equation}
With this last equation and equation (11), the flux ratio -
given by equation (14) - is fully determined.

\subsection{Time dependence of scattered emission}

For simplicity, let us assume that the emission from shell 1 is
constant and time independent: that it is a box function with some
finite duration.  The time dependence of the scattered emission will
be given by: (i) the time evolution of the electrons' optical depth,
$\tau_e$, (ii) the opening angle of shell 1 as seen by a co-moving
observer on shell 2, $\alpha_j''$, and (iii) the radius of shell 2,
$R_2(t_{scat})$.

The time dependence of $R_2(t_{scat})$ and $\alpha_j''$ is weak,
provided that the ejection time delay between shells, $\delta t$, is
small compared to $R_1$.  This is the case we are interested in,
since if $\delta t \sim R_1$, then $\delta t$ would be on the order
of hours or days. This scenario would invoke a very long lasting
activity of the central engine, a scenario that we don't want to
address in this paper.  The only time dependence of the scattered
emission will be given by the time evolution of $\tau_e$, which goes
as $\tau_e \propto R_2^{-2} \propto T^{-2}$.

\section{Primary and scattered emission relations}

\subsection{Time delay}

Let us assume that two photons are emitted from shell 1 at the same time.
Photon 1 travels directly to the observer located ahead of shell 1 and
arrives at time $T_p$ ($p$ stands for {\it primary}).  Photon 2 travels
in the opposite direction, scatters from shell 2
and then travels back to the same observer arriving at a later time
$T_s$ (where $s$ stands for {\it scattered}).
What is the time delay, $T_s - T_p$, between the arrival of these two photons?

If we are only interested, as a simplification, in the photons along
the line of sight, then this time delay will be given by equation (15).
We only need to multiply this expression by $2$, to obtain the full
time it takes for the photon to travel to shell 2 and then to travel back
to shell 1.  Then, the time delay is

\begin{displaymath}
T_s - T_p = 2 \frac{R_1(t)-R_2(t)}{1 + \beta_2}.
\end{displaymath}

Further simplification yields

\begin{equation}
T_s - T_p = R_1 \biggl(1 - \frac{\beta_2}{\beta_1}\biggr) + \beta_2 \delta t.
\end{equation}

\subsection{Ratio of frequencies}

In this section we will determine the relation between the frequency
of a photon emitted from shell 1, $\nu$, and a photon emitted from
shell 1 and then scattered by shell 2, $\nu_s$ (both quantities
measured in the laboratory frame).  For this, we will only consider
the photons traveling along the direct line of sight between the
central engine and the observer.

In the shell 1 co-moving frame, a photon is emitted from shell 1 with
frequency $\nu'$.  In the laboratory frame, this frequency is measured
as

\begin{displaymath}
\nu = \nu' \Gamma_1 (1 + \beta_1 \cos \Theta'),
\end{displaymath}
where $\Theta'$ is the angle between the photon's direction and
the $x'$-axis direction of the shell 1 co-moving frame. Note that the
$x'$-axis of frame $K'$ is always directed to the moving direction
of frame $K'$ relative to another inertial frame ($K$ or $K''$).
Since the photon is emitted to the observer ahead of shell 1,
$\Theta' = 0$, and the previous equation becomes

\begin{equation}
\nu = \nu' \Gamma_1 (1 + \beta_1).
\end{equation}

Let us now consider a photon emitted in shell 1 and directed to
shell 2. The frequency of this photon in the shell 1 co-moving frame
is $\nu'$, and in the shell 2 co-moving frame is given by $\nu''$.
They are related by

\begin{displaymath}
\nu'' = \nu' \Gamma''_1 (1 + \beta''_1 \cos \Theta''),
\end{displaymath}
where $\Gamma''_1$ ($\beta''_1 $) is the LF (velocity) of the shell
1 as measured in the shell 2 co-moving frame. These quantities can be
expressed in terms of quantities in the laboratory frame as

\begin{displaymath}
\Gamma''_1 = \Gamma_1 \Gamma_2 (1-\beta_1 \beta_2), \qquad
\beta''_1 = \frac{|\beta_1 - \beta_2|}{1-\beta_1 \beta_2 }.
\end{displaymath}
If we assume that $\Gamma_1 > \Gamma_2$ (or $\Gamma_1 < \Gamma_2$), in the
co-moving frame of shell 2 it will seem that shell 1 is moving away
from shell 2 (moving towards shell 2), so that $\Theta''=\pi$
($\Theta''=0$).  Using the last 3 equations, we can determine

\begin{equation}
\nu'' = \nu' \Gamma_1 \Gamma_2 (1 - \beta_1)(1 + \beta_2),
\end{equation} which holds for both assumptions.  The photon will be scattered by shell 2
by Thomson scattering (there will be no frequency change in the co-moving frame of
shell 2) and then will travel towards
the observer.  The scattered frequency can be obtained with

\begin{displaymath}
\nu_s = \nu'' \Gamma_2 (1 + \beta_2 \cos \Theta'').
\end{displaymath}
Setting $\Theta''\approx 0$, since the photon moves towards the observer,
yields

\begin{equation}
\nu_s = \nu'' \Gamma_2 (1 + \beta_2)
\end{equation}

Finally, using equations (18)-(20) the ratio of the scattered
frequency to the primary frequency is

\begin{equation}
\frac{\nu_s}{\nu} = \biggl(\frac{\Gamma_2}{\Gamma_1}\biggr)^2
\biggl(\frac{1+\beta_2}{1+\beta_1}\biggr)^2 \approx
\biggl(\frac{\Gamma_2}{\Gamma_1}\biggr)^2,
\end{equation} using the fact that both shells travel close to the speed of light.
This is consistent with our earlier simple estimation. Therefore, a slower (faster)
shell 2 will lower (raise) the frequency of the primary photons.

\subsection{Ratio of observed durations}

Let us assume that shell 1 emits for a finite duration of time, $\Delta t$
(in the lab frame).  An observer located in front of the shell will detect that the
radiation from shell 1 lasts for $\Delta T_p$, given by
$\Delta T_p = \Delta t (1-\beta_1)$.  The radiation from shell 1 will also travel
back to shell 2 and will get scattered, giving a scattered radiation
duration of $\Delta T_s$ in the observer frame. If the first photon from shell 1
is emitted at time $t$ and the last one at $t + \Delta t$ (in the lab frame),
then we can use the time delay equation (17) to find the time delay between the first
primary photon and the first scattered photon, and the time delay between the last
primary photon and the last scattered photon, respectively.
Subtracting these two expressions, we find that

\begin{equation}
\frac{\Delta T_s}{\Delta T_p} = \frac{1-\beta_2}{1-\beta_1}
\approx \biggl(  \frac{\Gamma_1}{\Gamma_2}\biggr)^2,
\end{equation} which means that the observed duration of the scattered
emission will be stretched (shortened) by a factor of
$\Big(\frac{\Gamma_1}{\Gamma_2}\Big)^2$ for a slower (faster) shell 2.

\begin{figure*}
\centerline{\hbox{\includegraphics[width=9cm,angle=270]{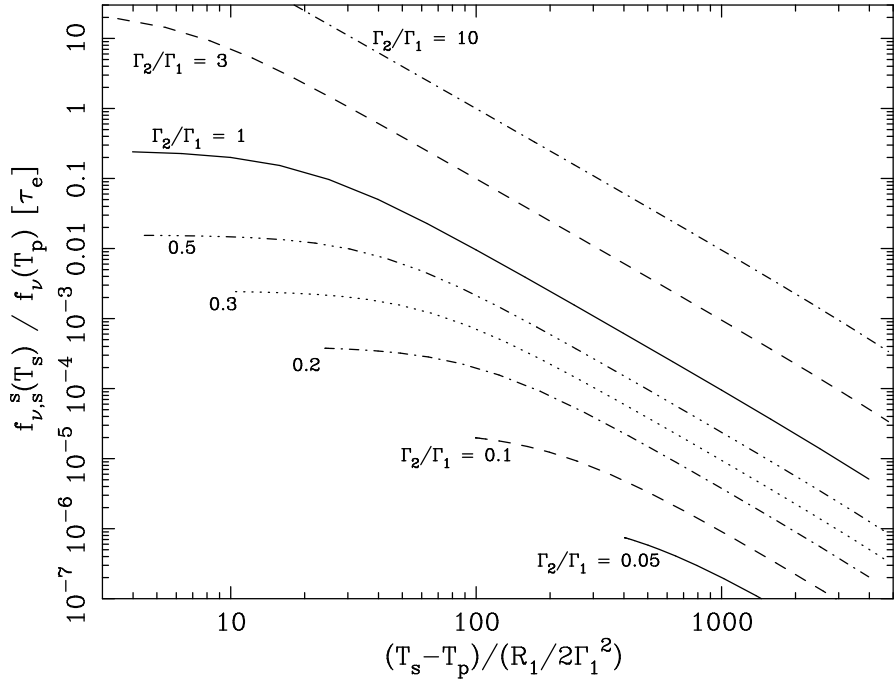}}}
\caption {The ratio of the flux of the scattered emission to the
flux of the primary emission (measured at their respective arrival
times and photon energies) as a function of the observed time delay
between these two emission components in units of $\frac{R_1}{2\Gamma_1^2}$,
assuming shell 1 LF $\Gamma_1$ = 100 and shell opening angle $\theta_j$ = 0.1.
The unknown shell 2 optical depth $\tau_e$ is put on the y-axis as
an ``unit'' of the flux density ratio. }
\end{figure*}

\section{Results}

\subsection{Ratio between scattered and primary fluxes}

We want to write equation (14), the ratio of the fluxes, in such a
way that we can easily use the available observations to test our
theory.

We can use equation (17) to solve for $\delta t$, the time delay between
the ejection of the two shells.  This expression is then substituted into
equation (11) to get $\sin^2{\alpha_j''}$ and into equation (16) to get
the radius ratio.  Finally, we obtain the flux ratio in terms of the time
delay divided by $\frac{R_1}{2\Gamma_1^2}$ as
follows:

\begin{displaymath}
\frac{f^{s}_{\nu,s}(T_s)}{f_{\nu}(T_p)}= \frac{1}{4} \tau_e
\biggl\{1 + \biggl[\frac{\Gamma_2 }{2 \Gamma_1^2 \theta_j}
\frac{(T_s-T_p)}{R_1/2\Gamma_1^2}\biggr]^2 \biggr\}^{-1}
\end{displaymath}
\begin{equation} \quad \quad \quad \quad \quad \times
\biggl[\frac{1}{2}+\frac{\beta_2}{2
\beta_1}-\frac{1}{4\Gamma_1^2}\frac{(T_s -
T_p)}{R_1/2\Gamma_1^2}\biggr]^2 \biggl(
\frac{\Gamma_2}{\Gamma_1}\biggr)^4.
\end{equation}

We then plot this ratio of two flux densities as a function of
the $\frac{R_1}{2\Gamma_1^2}$-normalized time delay (the choice of
this normalization will become evident on the next sections) for
various values of $\Gamma_2/\Gamma_1$ in Figure 2. For plotting
purposes we assume $\Gamma_1=100$ and $\theta_j=0.1$.
Since the exact value for $\tau_e$ is unknown, we choose to scale
$\tau_e$ into the y-axis of the figure, as an ``unit'' of the flux
density ratio.

Looking at figure 2, we can observe that the theoretical
flux ratio curves have two regions (this is very noticeable
when $\Gamma_2 < \Gamma_1$). In region I the flux
ratio is flat, and in region II the flux ratio is proportional
to the square of the inverse of the normalized time delay.  If we
inspect equation (23), we can separate its two regions by:

\begin{displaymath}
\frac{f^{s}_{\nu,s}(T_s)}{f_{\nu}(T_p)} = \end{displaymath}
\begin{displaymath}  \left\{
\begin{array}{ll} \frac{\tau_e}{4}\biggl(\frac{\Gamma_2}{\Gamma_1}
\biggr)^4,  &
\textrm{if $ \frac{T_s-T_p}{R_1 / 2\Gamma_1^2} \leq \frac{2 \Gamma_1^2 \theta_j}{\Gamma_2}$ \quad (Region I),}\\
\tau_e \Gamma_2^2 \theta_j^2 \biggl(  \frac{T_s-T_p}{R_1 / 2\Gamma_1^2}    \biggr)^{-2},  &
\textrm{if $ \frac{T_s-T_p}{R_1 / 2\Gamma_1^2} > \frac{2 \Gamma_1^2 \theta_j}{\Gamma_2}$ \quad (Region II).}\\
 \end{array} \right.
\end{displaymath}
These two regions will be used in our applications
section. For now, they just provide a simpler theoretical model.
Notice that the $\Gamma_2 > \Gamma_1$ curves are dominated
by region II, while the $\Gamma_2 < \Gamma_1$ curves have a combination
of both regions. For the latter, the maximum scattered flux is given by
region I.

These results show that the scattered flux from a slower shell is small, falls
at a lower energy than the primary photon's energy and its total
duration is larger than that of the primary emission.  If we have a
faster shell, then the scattered flux from it could be either larger or
smaller than the primary flux, depending on its ejection time delay,
$\delta t$.  If $\delta t$ is larger (smaller) than the total observed
duration of the primary emission, then the scattered emission would
appear at late times (would be part of the primary emission).  In any
case, the energy of the scattered photons would be larger than that of the
primary ones, and the total duration of the scattered emission would be
smaller than that of the primary emission, so that the scattered emission
would appear as a short bright flash.

\subsection{Hot shell 2}

In the last sections we assumed {\it Thomson scattering},
but we should also look at the possibility that shell 2 could be
hot, so the scattering mechanism at work would be {\it inverse Compton}.
Shell 2 electrons could be hot in the following scenario.
Consider that shell 2 is ejected after shell 1 and undergoes particle
heating, by either internal shocks or magnetic dissipation, at a
radius smaller than the radius where shell 1 produces $\gamma$-ray
photons. Shell 2 would experience adiabatic expansion and would cool,
but the electrons could still be hot by the time that the $\gamma$-ray
photons from shell 1 reach them. The shell 2 electrons might also cool
via radiation, but the chances still exist that the radiative cooling is
very inefficient, for instance, when the radiation mechanism is
synchrotron-self-inverse-Compton instead of pure synchrotron for the same
observed typical photon energy, so that the electron cooling time could be
comparable to the delay between the electron heating and the scattering.

The main difference in the formulas previously derived
will be in the ratio of the frequencies of the primary and the
scattered emission.  Equation (21) is modified to include the inverse Compton
boost to the photon energy:
\begin{equation}
\frac{\nu_s}{\nu} = \biggl(\frac{\Gamma_2}{\Gamma_1}\biggr)^2 \gamma_e^2,
\end{equation}
where $\gamma_e$ is the electrons' thermal Lorentz factor.
The theoretical flux density ratio previously derived will not be
changed in this new scenario.

Let us now calculate the isotropically equivalent total energy
in the shell 2 hot electrons.  For this, we need to calculate the
isotropically equivalent total number of electrons in the shell from
their optical depth

\begin{displaymath}
N_e = 4 \pi R_2^2 \frac{\tau_e}{\sigma_T}.
\end{displaymath}
In the present
scattering scenario, we have $R_2 \approx R_1$ (we'll prove
this on section \S7.2.2).  $R_1$ can be estimated from
the $\gamma$-ray variability time scale: $\frac{R_1}{2\Gamma_1^2}$
(explained in detail on \S7.1.1).  Then the total energy in
the hot electrons of shell 2 is given by

\begin{equation}
 E_e = N_e m_e c^2 \Gamma_2 \gamma_e =
16\pi\biggl(\frac{\Gamma_2}{\Gamma_1}\biggr)\Gamma_1^5\gamma_e m_e c^4
\biggl(\frac{R_1}{2\Gamma_1^2c}\biggr)^2\frac{\tau_e}{\sigma_T},
\end{equation} where $c$ is the speed of light.

\begin{table*}
\caption{Our sample of 10 GRBs which show clearly an X-ray shallower
decay component and the relevant data. Basic data are from O'Brien
et al. (2006).}
\begin{center}
\begin{minipage}{\textwidth}
\begin{tabular}{ccccccccccc}
\hline
GRB & 050315 & 050319 & 050713A & 050713B & 050714B & 050803 &
050814 & 050819 & 050822 & 050915B \\
\hline \hline

$F_{BAT}$
\symbolfootnote[1] {Mean BAT flux.}  & \raisebox{-2.0ex}{3.2} &
\raisebox{-2.0ex}{0.5} & \raisebox{-2.0ex}{3.8} &
\raisebox{-2.0ex}{3.2} & \raisebox{-2.0ex}{1.1} &
\raisebox{-2.0ex}{2.4} & \raisebox{-2.0ex}{1.2} &
\raisebox{-2.0ex}{0.9} & \raisebox{-2.0ex}{2.5} & \raisebox{-2.0ex}{8.6} \\
(10$^{-8}$erg cm$^{-2}$ s$^{-1}$) & & & & & & & & & & \\
\hline

$\beta_{BAT}$
\symbolfootnote[2] {BAT spectral index.} & 1.2 & 1 & 0.6 & 0.5 & 2 &
0.5 & 1 & 1.6 & 1.5 & 1 \\
\hline

$F_{\nu=100keV}(T_{90})$
\symbolfootnote[3] {BAT flux density at 100 keV at the end of the
$\gamma$-ray emission.} & \raisebox{-2.0ex}{4} &
\raisebox{-2.0ex}{0.8} & \raisebox{-2.0ex}{12} &
\raisebox{-2.0ex}{6} & \raisebox{-2.0ex}{0.6} &
\raisebox{-2.0ex}{4.8} & \raisebox{-2.0ex}{2} & \raisebox{-2.0ex}{1}
& \raisebox{-2.0ex}{3.2} & \raisebox{-2.0ex}{16} \\
(10 $\mu$Jy) & & & & & & & & & & \\
\hline

$F_{\nu=10keV}(T_{90})$
\symbolfootnote[4] {BAT flux density at 10 keV at the end of the
$\gamma$-ray emission.} & \raisebox{-2.0ex}{63} &
\raisebox{-2.0ex}{8} & \raisebox{-2.0ex}{48} & \raisebox{-2.0ex}{19}
& \raisebox{-2.0ex}{60} & \raisebox{-2.0ex}{15} &
\raisebox{-2.0ex}{20} & \raisebox{-2.0ex}{40} &
\raisebox{-2.0ex}{101} & \raisebox{-2.0ex}{160} \\
(10 $\mu$Jy) & & & & & & & & & & \\
\hline

$T_{90}$
\symbolfootnote[5] {Duration of the $\gamma$-ray emission.} &
\raisebox{-2.0ex}{96} & \raisebox{-2.0ex}{150} &
\raisebox{-2.0ex}{130} & \raisebox{-2.0ex}{130} &
\raisebox{-2.0ex}{50} & \raisebox{-2.0ex}{90} &
\raisebox{-2.0ex}{144} & \raisebox{-2.0ex}{36} &
\raisebox{-2.0ex}{105} & \raisebox{-2.0ex}{40} \\
(s) & & & & & & & & & & \\
\hline

$\frac{R_1}{2\Gamma_1^2}$
\symbolfootnote[6] {Simply $T_{90}$ for those bursts with one smooth
or two overlapped pulses; for those ``spiky'' bursts, i.e., those
with multiple, separated pulses, we use the duration of the last
pulse.} & \raisebox{-2.0ex}{30} & \raisebox{-2.0ex}{40} &
\raisebox{-2.0ex}{20} & \raisebox{-2.0ex}{130} &
\raisebox{-2.0ex}{50} & \raisebox{-2.0ex}{90} &
\raisebox{-2.0ex}{144} & \raisebox{-2.0ex}{36} &
\raisebox{-2.0ex}{30} & \raisebox{-2.0ex}{40} \\
(s) & & & & & & & & & & \\
\hline

$t_{end}$
\symbolfootnote[7] {Ending time of the X-ray shallow decay.} &
\raisebox{-2.0ex}{1} & \raisebox{-2.0ex}{3.2} & \raisebox{-2.0ex}{1}
& \raisebox{-2.0ex}{4} & \raisebox{-2.0ex}{5} & \raisebox{-2.0ex}{2}
& \raisebox{-2.0ex}{$\ge$ 6} & \raisebox{-2.0ex}{2} &
\raisebox{-2.0ex}{1.3} & \raisebox{-2.0ex}{5} \\
($10^4$ s) & & & & & & & & & & \\
\hline

$F_{XRT}(t_{end})$
\symbolfootnote[8] {XRT flux at the end of the shallow decay.} &
\raisebox{-2.0ex}{0.8} & \raisebox{-2.0ex}{0.8} &
\raisebox{-2.0ex}{1.2} & \raisebox{-2.0ex}{1} &
\raisebox{-2.0ex}{0.5} & \raisebox{-2.0ex}{1} &
\raisebox{-2.0ex}{$\le$ 0.05} & \raisebox{-2.0ex}{0.04} &
\raisebox{-2.0ex}{0.7} & \raisebox{-2.0ex}{0.09} \\
($10^{-11}$ erg cm$^{-2}$ s$^{-1}$) & & & & & & & & & & \\
\hline

$\beta_X$
\symbolfootnote[9] {XRT spectral index.} & 1.5 & 2 & 1.3 & 0.7 & 4.5
& 0.7 & 1.1 & 1.2 & 1.6 & 1.5 \\
\hline

$F_{\nu=1 keV}(t_{end})$
\symbolfootnote[10] {XRT flux density at 1 keV at the end of the
shallow decay.} & \raisebox{-2.0ex}{1.2} & \raisebox{-2.0ex}{1.1} &
\raisebox{-2.0ex}{2.4} & \raisebox{-2.0ex}{1} &
\raisebox{-2.0ex}{0.16} & \raisebox{-2.0ex}{1} &
\raisebox{-2.0ex}{$\le$ 0.06} & \raisebox{-2.0ex}{0.04}
& \raisebox{-2.0ex}{1.2} & \raisebox{-2.0ex}{0.1} \\
($\mu$Jy) & & & & & & & & & & \\
\hline

$\frac{t_{end}}{R_1/(2\Gamma_1^2)}$
\symbolfootnote[11] {Observed time delay between the primary and the
scattered emissions in units of $\frac{R_1}{2\Gamma_1^2}$. See \S
7.1.2.} & 330 & 800 & 500 & 308 & 1000 & 222 & $\ge$ 417 & 556 & 433
& 1250 \\
\hline

$\frac{F_{\nu=1keV}(t_{end})}{F_{\nu=100keV}(T_{90})}$ & 0.03 &
0.137 & 0.02 & 0.017 & 0.026 & 0.021 & $\le$ 0.003 & 0.004
& 0.037 & 0.0006 \\
\hline

$\frac{F_{\nu=1keV}(t_{end})}{F_{\nu=10keV}(T_{90})}$ & 0.0019 &
0.0138 & 0.005 & 0.0052 & 0.0003 & 0.0066 & $\le$ 0.0003
& 0.0001 & 0.0012 & 0.00006 \\
\hline

\end{tabular}
\end{minipage}
\end{center}
\end{table*}

\section{Application to the shallow decay component in GRB early X-ray afterglows}

In this section we compare the $\gamma$-ray burst and X-ray afterglow
data with the results of the last section to find out if the shallow decay
in X-ray ``canonical'' afterglow light curves can be due to the scattered
emission from the shell(s) following the $\gamma$-ray shell. We first
consider simply the Thomson scattering mechanism, then we turn to
consider the inverse Compton scattering where the electrons
in the scattering shell have highly relativistic thermal energy.

\subsection{Thomson scattering}

\subsubsection{Data set}

From a {\it Swift} GRB early X-ray afterglow catalog presented in
O'Brien et al (2006),
we choose a sample of 10 bursts all of which show clearly a
``canonical'' behavior that includes a shallow decay component. We
apply our simple scenario to these bursts, and assume that: (1) the
last $\gamma$-ray photon that was emitted from shell 1 traveled
directly to the observer (primary emission); (2) at the same time
and from the same site on shell 1, another photon traveled to shell
2, was scattered, and eventually became the last X-ray photon of the
shallow decay phase (scattered emission). Therefore, we will use the
ratio of the flux density at X-ray energy at the end of the shallow
decay and the flux density at $\gamma$-ray energy at the end of the
$\gamma$-ray emission.  Theoretically, this ratio should fit our
equation (14) if the shallow decay were to have its origin in
the scattered emission scenario.

For the $\gamma$-ray photons, the catalog in O'Brien et al. (2006) only
gives a mean BAT flux. We use this mean flux as an approximation to
the flux at the end of $\gamma$-ray emission; the flux density at a
specific photon energy is obtained using the BAT spectral index
$\beta_{BAT}$ ($f_{\nu} \propto \nu^{-\beta_{BAT}}$). For the X-ray
photons, we use the XRT flux at the end of the shallow decay and the
X-ray spectra index $\beta_X$ ($f_{\nu} \propto \nu^{-\beta_X}$) to
obtain the flux density at a specific photon energy.

In our theory $R_1$ is an unknown parameter, but we should be able
to extract it from the available data: $\frac{R_1}{2\Gamma_1^2}$ is
the $\gamma$-ray burst duration $T_{90}$ for FREDs (fast rise and
exponential decay) - those bursts
whose light curves are made of one smooth or two overlapped pulses;
and for bursts with multiple spikes in light curves,
$\frac{R_1}{2\Gamma_1^2} < T_{90}$. Note that
$\frac{R_1}{2\Gamma_1^2}$ is equal to the curvature time scale - the
delay between two photon's arrival times, one emitted from shell 1's
visible edge and the other from the center of shell 1's visible
region. Thus, the $\gamma$-ray variability time scale, if we assume
it is mainly determined by the curvature time scale, would be a good
approximation for $\frac{R_1}{2\Gamma_1^2}$. Therefore, we look up
the $\gamma$-ray light curves from the Swift
archive\footnote{http://heasarc.gsfc.nasa.gov/docs/swift/archive/grb\_table/};
for FREDs, $\frac{R_1}{2\Gamma_1^2}$ is simply $T_{90}$, but for
those spiky bursts, we use the duration of the {\it last} pulse. The
data obtained for our sample of 10 bursts has been organized in
Table 1.

\subsubsection{Comparison between observations and theory predictions}

For the X-ray shallow decay $\Gamma_2/\Gamma_1=1$ is a limiting case,
since for $\Gamma_2/\Gamma_1>1$: (i) the scattered emission from the
$\gamma$-rays would fall at a higher energy, not in the X-rays,
according to equation (21),  and (ii)  would have a smaller duration
than the $\gamma$-rays duration according to equation (22), which is
not what it is observed - the shallow decay in X-ray light curves
typically extends up to $10^4$ s.

In Figure 3, we plot our 10 GRBs sample data and the results of our
theoretical calculations for two cases.  For the first case, the 10
data points use the XRT flux at 1 keV (scattered emission) and the
BAT flux at 100 keV (primary emission), which corresponds to
$\Gamma_2/\Gamma_1=0.1$ according to equation (21). For the time
delay, we use $(T_s-T_p) \approx t_{end}$, where $t_{end}$ is the
end time of the shallow decay - we do this because $T_{90} \ll
t_{end}$. For $\frac{R_1}{2\Gamma_1^2}$, we use the method described
in the last subsection. These values are also listed in Table 1.

In the same figure, for the second case, the 10 data points use the
XRT flux at 1 keV (scattered emission) and the BAT flux at 10 keV (primary
emission). This corresponds to $\Gamma_2/\Gamma_1=1/\sqrt{10}\approx
0.3$, according to equation (21).  We consider this case to see if
the shallow decay might be produced
by the scattering of the low energy tail of the $\gamma$-ray
emission. Since in this case shell 2 is faster than in the
first case described above, the time delay in the ejection of the
shells will be larger.

The normalized time delay for the sample has a range of
$2\times10^2 < \frac{T_s-T_p}{R_1/2\Gamma_1^2} < 2\times10^3$.
For fiducial parameter values $\Gamma_1$= 100, $\Gamma_2$= 10 and
$\theta_j$= 0.1, using equations (11) and (17), this range has a
constraint on $\alpha_j''$: $0.1<\tan\alpha_j''<1$. Since this angle
is not very small, the approximation of the homogeneous incident
flux on shell 2 that was used in deriving the scattered flux formula
in \S 3.1 will slightly overestimate the observed scattered flux
(see discussion in footnote 1).

\subsubsection{Results}

Figure 3 shows that the observed flux ratios of the sample are
generally $(10^3-10^4)\tau_e^{-1}$ times larger than the theoretical
expectation. The discrepancy would be a factor of $\sim 10^4-10^5$
for a modest value of $\tau_e \sim 0.1$. This indicates that the
emission of the shallow decay is too luminous to be interpreted
simply by the scattering within the two shell scenario.

The same figure shows a smaller discrepancy between the sample data
and the theoretical curve for a case where shell 2 is slightly faster
but still not exceeding the $\gamma$-ray shell's speed.
Using a modest value of $\tau_e \sim 0.1$ would make this discrepancy
to be $\sim 10^2$.

\begin{figure*}
\centerline{\hbox{\includegraphics[width=9cm,angle=270]{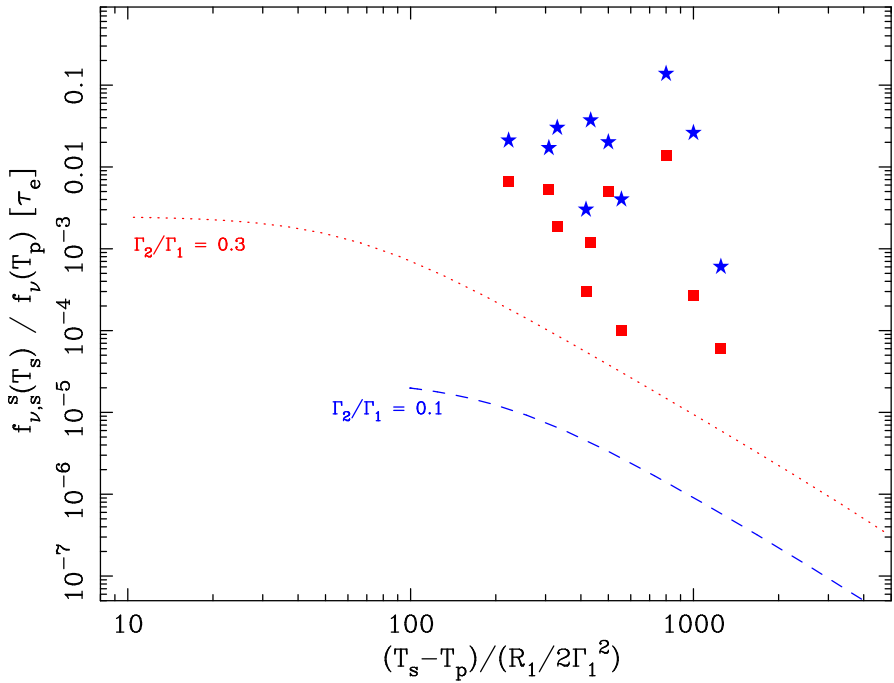}}}
\caption {Similar to Figure 1 but with the GRB sample data added.
The stars (squares) are our GRB sample data using the ratio between
the XRT flux density at the end of the shallow X-ray decay at 1 keV
and the mean BAT flux density at 100 keV (10 keV). The dashed (dotted)
line is the result of our theoretical calculation.}
\end{figure*}

\subsection{Inverse Compton scattering in shell 2}

\subsubsection{Comparison between observations and theory predictions}

To compare the fluxes of the X-ray shallow decay with the theoretical
expectation of the inverse Compton scattering scenario, we fix the scattered
frequency, $\nu_s$ = 1 keV. With equation (24) we can determine the
frequency of the primary emission. The flux ratio data points in Figure 3
have to be changed because the frequency ratio is changed.
The flux in the $\gamma$-ray band follows $f_{\nu}
\propto \nu^{-\beta_{BAT}}$. With this, we can now determine that
the data points in Figure 2 will be multiplied by a factor of
$[\frac{\nu_{\gamma}}{\nu_X}(\frac{\Gamma_2}{\Gamma_1})^2\gamma_e^{2}]^{-\beta_{BAT}}$
to account for the inverse Compton scattering effect, where $\nu_{\gamma}$
and $\nu_X$ are the $\gamma$-ray and X-ray photon frequencies, respectively,
at which the flux densities are used in the data points of Figure 3.

Since $\gamma_e$ is an unknown
quantity we cannot determine where these new points would lie on a
plot analogous to Figure 3.  For this reason, we ask: what
is the value of $\gamma_e$ necessary to lower all data points to the
theoretically expected curve?  To answer, let us fix the value of the
Lorentz Factor ratio, $\Gamma_2/\Gamma_1 = 0.1$ -  this is a reasonable
value according to equation (22), since the $\gamma$-ray duration usually
lasts $\sim 10^2$ s and the shallow decay extends to $\sim 10^4$ s.

The required values for $\gamma_e$ are presented in Table 2. We present
two different values for each burst.  We use the BAT spectral index
(subscript $c$) and an average between the BAT and XRT spectral indices
(subscript $e$), respectively.  We do this because we are extrapolating
the $\gamma$-ray flux to energies below the observed BAT band and it is
unknown if the BAT spectrum will behave as a single power law in this region.

\subsubsection{Electrons' energy in shell 2}

We calculate $E_e$ for each burst in our sample using
equation (25).  The results are presented in Table 2, adopting $\Gamma_1=10^2$,
$\frac{\Gamma_2}{\Gamma_1}=0.1$ and $\tau_e$ = 0.1. Note that we have
corrected $\frac{R_1}{2\Gamma_1^2}$ for the cosmological time dilation
effect. Two values of energies from the two values of
$\gamma_e$ obtained in the last subsection are given in the table.

Note that from equations
(16) and (17) one can see that $\frac{R_2}{R_1}=
\frac{1}{2}+\frac{\beta_2}{2\beta_1}-\frac{T_s-T_p}{2R_1} \simeq 1-
\frac{1}{4\Gamma_1^2}\frac{(T_s-T_p)}{R_1/2\Gamma_1^2}$. In our data
sample $\frac{T_s-T_p}{R_1/2\Gamma_1^2}\leq 10^3$. Therefore,
the approximation made in the derivation of equation (25), that $R_2
\approx R_1$ for $\Gamma_1 \sim 10^2$, is valid.

\begin{table*}           
\caption{Calculated values for the required $\gamma_e$ and total
electron energy of shell 2 to obtain the theoretical expected flux
ratio. The calculations were done with two different spectral
indices: the BAT spectral index and the average between the BAT
spectral index and the XRT spectral index.}
\begin{center}
\begin{minipage}{\textwidth}
\begin{tabular}{ccccccc}
\hline
 \raisebox{-3.0ex}{GRB}
 & \raisebox{-3.0ex}{Redshift
\symbolfootnote[1] {References for known redshifts: GRB 050315:
Kelson \& Berger (2005); GRB 050319: Fynbo et al. (2005); GRB
050814: Jakobsson et al. (2006). For bursts without measured
redshift, we use the mean redshift $z$= 2.8 of the Swift GRB
redshift distribution (Jakobsson et al. 2006).}} &
$\frac{R_1}{2\Gamma_1^2}$ \symbolfootnote[2] {Cosmological time
dilation
  corrected curvature variability time scale; it is equal to the
  $\frac{R_1}{2\Gamma_1^2}$ in Table 1 divided by $(1+z)$.}
& $\gamma_e$ \symbolfootnote[3] {Required shell 2 electron thermal
LF using
  $\beta_{BAT}$ as the spectral index.}
& $E_e$ \symbolfootnote[4] {Isotropically equivalent total energy of
electrons with $\gamma_e$ calculated in the last previous column for
$\tau_e$ = 0.1.} & $\gamma_e$ \symbolfootnote[5] {Required shell 2
electron thermal LF using
  $(\frac{\beta_{BAT}+\beta_X}{2})$ as the spectral index.}
& $E_e$ \symbolfootnote[6] {Isotropically equivalent total energy of
electrons with $\gamma_e$ calculated in the last previous column for
$\tau_e$ = 0.1.}
 \\
 & & (s) & [$\tau_e^{-\frac{1}{2\beta}}$] &  (erg) &
[$\tau_e^{-\frac{1}{2\beta}}$] & (erg)
 \\
\hline \hline
050315 & 1.95 & 10 & 34 & 5 $\times 10^{52}$ & 23 & 3 $\times 10^{52}$\\
050319 & 3.24 & 9.4 & 316 & 5 $\times 10^{53}$ & 46 & 5 $\times 10^{52}$\\
050713A & & 5.3 & 1364 & 1.4 $\times 10^{54}$ & 95 & 5 $\times 10^{52}$\\
050713B & & 34 & 2404 & 1.5 $\times 10^{56}$ & 657 & 3 $\times 10^{55}$\\
050714B & & 13 & 13 & 2 $\times 10^{52}$ & 5 & 7 $\times 10^{51}$\\
050803 & & 24 & 1931 & 6 $\times 10^{55}$ & 547 & 1 $\times 10^{55}$\\
050814 & 5.3 & 23 & 25 & 2 $\times 10^{53}$ & 22 & 2 $\times 10^{53}$\\
050819 & & 9.5 & 10 & 1 $\times 10^{52}$ & 13 & 1.5 $\times 10^{52}$\\
050822 & & 8 & 21 & 1.6 $\times 10^{52}$ & 19 & 1.4 $\times 10^{52}$\\
050915B & & 10.5 & 33 & 6 $\times 10^{52}$ & 17 & 3 $\times 10^{52}$\\
\hline
\end{tabular}
\end{minipage}
\end{center}
\end{table*}

\subsubsection{Results}

Table 2 shows that the isotropically equivalent total energy carried
by the electrons of a hot shell 2 is large, $\sim 10^{52-56}$ erg.
If we take into account the cooling of the electrons via adiabatic
expansion and/or radiation, the initial total
energy when the electrons were just accelerated is even bigger. The
optical depth, $\tau_e$, would certainty decrease the total energy
but only by a small fraction. The prompt emission from those
electrons in shell 2 would arrive at about the same time as the
scattered emission, and the two emission components would have
similar durations because both durations are $\propto 1/\Gamma_2^2$.
Depending on the ejecta properties, e.g., the ratio of the shell 2
energy to the shell 1 energy, the shell 2 prompt emission might
dominate the scattered emission in the light curve. If that is true,
the shell 2 prompt emission might be a possible origin of the late
X-ray flares in bursts for which both the flares and the shallower
decay are present.

\section{Faster shell 2}

In the previous section we assumed a slower shell
2. What if shell 2 is faster than shell 1? Based on the formulae we
have, if $\Gamma_2 > \Gamma_1$, the scattered emission from
$\gamma$-rays would fall in higher energies, e.g., $\sim$ MeV, not
in the X-rays, and have a shorter duration
than the $\gamma$-rays.

It is shown in Figure 2 that, in the $\Gamma_2 > \Gamma_1$ cases,
the scattered emission is very bright
though it decreases with increasing time delay. If the scattered
emission spectrum mimics the power law form of the primary emission
spectrum, at some time delay significantly larger than the duration of
the primary emission, the lower-energy-extrapolated scattered flux is
close to or even brighter than the prompt $\gamma$-rays.
According to Figure 2,  for $\frac{\Gamma_2}{\Gamma_1}
\simeq3$ (corresponding to $\frac{\nu_s}{\nu}\simeq10$
according to equation (21)) and a
selected observed time delay $(T_s-T_p)/(\frac{R_1}{2\Gamma_1^2})
\simeq 100$ (which corresponds to an ejection delay $\delta t \sim 10^3$s),
$f_{\nu,s}^s(T_s) \approx 0.1 [\tau_e] f_{\nu}(T_p)$.
Extrapolating the scattered flux density from $\nu_s$ to $\nu$, we have
$f_{\nu}^s(T_s)=(\frac{\nu_s}{\nu})^{\beta_{BAT}}\times 0.1 [\tau_e]
f_{\nu}(T_p) \approx f_{\nu}(T_p) [\tau_e]$ for $\beta_{BAT} \approx$
1. For a smaller observed time delay, the flux is even greater.
That means we should have seen a lagged very short $\gamma$-ray
flash at $\sim 10^2 - 10^3$ seconds after the burst, provided that
$\tau_e \approx 1$. This case cannot happen
because the observations have never showed this feature. Even though the flux of
the scattered emission is smaller for a larger observed time delay, in order
for the delayed scattered $\gamma$-ray flash to indeed happen but below the BAT
flux limit, shell 2 must have an extremely large ejection time delay $\delta t\ge$
a few $\times 10^3$ seconds (cf. equation 17) which is very difficult to explain
in terms of the central engine activity. For $\tau_e \ll 1$, the
$\Gamma_2 > \Gamma_1$ case could have happened but the flux would be too small
to be detected.

One possible case of $\Gamma_2 > \Gamma_1$ is that the shell 2 ejection
delay $\delta t$ is small and shell 2 has moved very close to shell 1 when
the scattering happens so that the observed time delay of the scattered
emission is comparable or smaller than the duration of the primary emission.
The scattered emission in the BAT energy range will become part of the observed
prompt $\gamma$-ray emissions in time. Future high energy ($\ge$ MeV) observations
(e.g., by {\it GLAST}) during the burst may be able to determine the existence of
this case.

\section{X-Ray dim or dark GRBs}

In \S 7, our calculations show that the scattered flux is much lower
than the observed X-ray shallow decay flux. If the scattering indeed
happens, then the resultant emission must have been buried in the
shallow decay. Thus, in order for the scattered emission to be
detected, not only the shallow X-ray decay component must be absent,
but also the normal external forward shock component must be
extremely weak or absent. In rare cases Swift does observe X-ray
afterglows without a shallow decay and without the normal forward
shock decay ($\alpha_X \sim 1 - 1.3$), which we call {\it X-ray dim
GRBs}. They show instead a very steep flux decay ($\alpha_X \ge 3$)
and are thought to be located in extremely low density regions.  The
steep decay component can be explained by the large angle emission
(Kumar \& Panaitescu 2000). GRB 050421 and GRB 051210 are two
examples (Godet et al. 2006; La Parola et al. 2006). Note that GRB
051210 is of the short burst class which, according to its compact
binary progenitor model, is more probable to occur in a low density
environment.  At late times, however, when the large angle emission
is low enough, neither burst shows any sign of re-brightening atop
the steep power-law decay.

Another case of interest is the existence of
{\it X-ray dark GRBs}:
short bursts (GRB 050906 and GRB 050925)
that show no X-ray afterglow
detection, only upper limits at
$\approx$ 100 s (Nakar 2007).  It is also believed that
these bursts took place in extremely low density environments.

X-ray dim and dark GRBs provide a great opportunity to put
into work the theory presented in this paper. If we
assume that there is a late ejecta behind the $\gamma$-ray
producing source, then we can try to constrain its LF by
looking at these bursts. Since their afterglow emission is
so weak (or not present), we can ask the question:
what are the constraints on the LF of the late ejecta,
so that the scattered emission is present in X-ray dim and
dark GRBs, but doesn't exceed the actual flux observations
or upper limits? We will devote this section to answer this
question.

We'll start by using the data from {\it Swift}'s
BAT and XRT observations as follows:

\begin{displaymath}
\frac{f^{s}_{\nu,s}(T_s)}{f_{\nu}(T_p)}
\leq \frac{f^{XRT}_{\nu,s}(T_s)}{f^{BAT}_{\nu}(T_{90})}
= 0.01 \frac{f^{XRT}_{1keV}(T_s)}{f^{BAT}_{100keV}(T_{90})}
\biggl(\frac{\Gamma_1}{\Gamma_2} \biggr)^2,
\end{displaymath}
where we have selected $\nu$ = 100 keV, but we have made no
assumption on the value of $\nu_s$ only that it should
obey equation (21).  We have also assumed $\beta_{X}=1$,
which is consistent with {\it Swift} observations.

With the previous inequality and using the two previously defined
regions of the flux ratio (\S6), we can find constraints
on the LF ratio.  In region I, we find:

\begin{equation}
\frac{\Gamma_2}{\Gamma_1} \leq
\min\bigg\{20 \biggl(\frac{T_s-T_p}{R_1 / 2\Gamma_1^2}\biggr)^{-1}
\theta_{j,-1} \Gamma_{1,2},
1.26 f_{x \gamma}^{1/6} \tau_e^{-1/6}\bigg\}
\end{equation} and in region II:

\begin{displaymath}
20 \biggl(\frac{T_s-T_p}{R_1 / 2\Gamma_1^2}\biggr)^{-1}
\theta_{j,-1} \Gamma_{1,2} < \frac{\Gamma_2}{\Gamma_1}
\leq \end{displaymath}
\begin{equation}
\quad \quad \quad \quad \quad 0.1 f_{x \gamma}^{1/4}
\biggl(\frac{T_s-T_p}{R_1 / 2\Gamma_1^2}\biggr)^{1/2}
\theta_{j,-1}^{-1/2} \Gamma_{1,2}^{-1/2} \tau_e^{-1/4},
\end{equation} where we have used
$f_{x \gamma} \equiv \frac{f^{XRT}_{1keV}(T_s)}{f^{BAT}_{100keV}(T_{90})}$
and the convention $Q_x = Q/10^x$ has been adopted.

After presenting these last two conditions, we should
return to the physical picture. The initial assumption
we made is that the scattered emission should be present
independently of the burst's data and its region.  This
means that the LF of the late ejecta could be very small,
so that its contribution to the flux would be also minuscule.
Therefore, we should discard the lower limit on equation (27).
Now, we want to constrain the LF of the late ejecta from
above; we are interested in knowing what's its maximum.
The maximum value between the two upper
limits of equations (26) and (27) will be the best
and more conservative value to choose:

\begin{displaymath}
\frac{\Gamma_2}{\Gamma_1} \leq \max \left\{ \begin{array}{ll}
\min \left\{ \begin{array}{ll}
20 \biggl(\frac{T_s-T_p}{R_1 / 2\Gamma_1^2}\biggr)^{-1} \theta_{j,-1} \Gamma_{1,2} \\
1.26 f_{x \gamma}^{1/6} \tau_e^{-1/6} \\ \end{array} \right\}  \\
0.1 f_{x \gamma}^{1/4} \biggl(\frac{T_s-T_p}{R_1 / 2\Gamma_1^2}\biggr)^{1/2}
\theta_{j,-1}^{-1/2} \Gamma_{1,2}^{-1/2} \tau_e^{-1/4}\\
 \end{array} \right\}.
\end{displaymath}

Now we are ready to consider the 4 bursts mentioned at the
beginning of this section and obtain the constraints for the LF
of the late ejecta.

Using the last X-ray detection of GRB 051210 (La Parola et al. 2006)
and the upper limit of GRB 050421 (Godet et al. 2006), both at $\sim
10^3$ s, we can obtain the ratio between the XRT flux density at 1
keV (scattered emission) and the BAT flux density (Chincarini et al.
2007) at 100 keV (primary emission), $f_{x \gamma}$. Following the
procedure outlined in \S 7.1.1, the values for
$\frac{T_s-T_p}{R_1/2\Gamma_1^2}$ are also obtained\footnote{When
following \S 7.1.1 to determine the normalized time delay, there is
some uncertainty determining the width of the pulses since the data
shows statistical noise. For GRB 050421, two possible values for the
normalized time delay were obtained hence two constraints for the LF
ratio were derived.  The value reported here is the more
conservative one.}. We can do exactly the same for the {\it X-ray
dark short GRBs}, using the upper limits of GRB 050906 and GRB
050925 at $\approx$ 100 s (Pagani et al. 2005; Beardmore et al.
2005; Nakar 2007). With $f_{x \gamma}$ and the normalized time
delay, we find the following constraints:

\begin{displaymath}
\frac{\Gamma_2}{\Gamma_1}(\theta_{j,-1}\Gamma_{1,2})^{1/2} \tau_e^{1/4}
\lae \left\{ \begin{array}{ll}
0.15 &
\textrm{for GRB 050906 and 050925,}\\
0.48 &
\textrm{for GRB 050421,}\\
0.78 &
\textrm{for GRB 051210.}\\
 \end{array} \right.
\end{displaymath}

The upper limits for the {\it X-ray dark short GRBs} provide
the best constraints, since the upper limits in their X-ray
flux are very strict.
These results imply that, for $\tau_e=1$, $\Gamma_1=100$ and $\theta_j=0.1$,
the late ejecta is very slow  $\Gamma_2 \lae 15$, but it
could be faster if $\tau_e$ decreases.

\section{Summary and Conclusions}

We have investigated a scenario of photons scattering by electrons
within a relativistic outflow. The outflow is composed of discrete
shells. One front shell emits radiation, observed as the GRB's
prompt $\gamma$-ray photons. Some fraction of the radiation is
incident backwards to the shell(s) behind, and is scattered
isotropically in the local rest frame. The scattered emission
arrives at the observer at a later time, $T_s$, and at a different
photon energy, $\nu_s$, that are determined by
the LF ratio of the two
shells and the time delay of the ejection of the second shell.
We calculated the flux density ratio, i. e., the flux
density of the delayed scattered emission to that of the front
shell's primary emission, as a function of the normalized arrival
time delay and the assumed LF ratio.

The calculated flux density ratio are compared with observations
using a sample of {\it Swift} GRB X-ray afterglows which show a distinct
shallower decay component in their light curves, with the motivation
to see if the scattering scenario could be the origin of
the shallower decay. The results are negative. For
Thomson scattering, the flux density of the scattered emission is
about $10^{3-4} \tau_e^{-1}$ times lower than that of the shallower
decay component, where $\tau_e$ is the scattering shell's electron
optical depth.

We also consider the inverse Compton scattering scenario in which
the electrons in the scattering shell is hot. We find that, in order
for the scattered emission flux to be bright enough to match the
shallower decay component, the isotropic equivalent of the total
energy carried by the hot electrons is large, $\sim 10^{52-56}$ erg.
The prompt emission from the scattering shell appears at the same
time as the scattered emission and with a similar duration.

In the cases where shell 2 is faster than shell 1, when extrapolated
to the BAT energy band, the scattered flux can be as bright as the
emission from shell 1. The delay of the scattered emission is determined
by the ejection delay of shell 2. When the ejection delay of shell 2 is
much larger than the duration of the primary emission, the scattered
emission would appear as a late short $\gamma$-ray/MeV flash. For a small
ejection delay of shell 2, the scattered emission would become part of
the observed prompt emission. The fact that no late short $\gamma$-ray/MeV
flash is detected does not support the existence of a late faster shell.

Lastly, we study the possibility of detection of the scattering
emission in two X-ray dim GRBs - that only show a very steep flux
decay and do not show either a X-ray shallow decay nor the normal
forward shock component - and in two X-ray dark short GRBs - that
show no X-ray afterglow detection at $\sim$ 100 s. Assuming that
there is slower moving ejecta material behind the fast $\gamma$-ray
producing shell in these bursts, we find upper limits for the Lorentz
factor of the late slower material. More sensitive observations of
X-ray dark short GRBs could provide stronger constraints on the
presence and properties of slower moving material accompanying
the fast $\gamma$-ray jet in GRBs.

Almost simultaneous to the appearance of this paper, Panaitescu (2007)
presents a similar work on the scattering of the GRB early emission
photons by a late outflow. Both papers present the same physics and the
main formulae are consistent. The main differences are that (i) we have
considered the scattering within the relativistic outflow and the photons
to be scattered are the prompt $\gamma$-ray photons, whereas Panaitescu
(2007) considers the scattering of the afterglow forward shock photons
by a late relativistic outflow, and (ii) Panaitescu (2007) has considered
a faster second shell to be able to explain features like the ``shallow decay''
and the X-ray flares, whereas our focus is mostly on a slower second shell.

\section*{Acknowledgments}
This work is supported in part by grants from NSF (AST-0406878) and NASA
Swift-GI-program. RFS and RBD thank E. McMahon and J. Johnson for helpful
discussions and suggestions.



\renewcommand{\theequation}{A-\arabic{equation}}
\setcounter{equation}{0}  
\section*{APPENDIX: Approximation to the integrand function in calculation of
the incident flux on shell 2}  

Here we show that the integrand function in equation (6) for the
incident flux from shell 1 on point $P$ in shell 2 comoving frame is
insensitive to $\sin ^2\alpha''$ so that the integrand can be
taken out of the integral as a constant.  We will also show that the
integrand is of order unity.

Let us call the integrand function $F(\alpha, \theta)$. All
equations we have are
\begin{equation}
F(\alpha, \theta)=
\frac{(1+\beta_2\cos\alpha)^3}{[\beta_1+\cos(\alpha-\theta)][1+\beta_1\cos(\alpha-\theta)]^2}
\end{equation}
\begin{equation}
\tan\alpha''=
\frac{\theta}{\Gamma_2(1-\frac{\beta_2}{\beta_1}+\frac{\beta_2\delta
t}{R_1})}
\end{equation}
\begin{equation}
\tan\alpha''= \frac{\sin\alpha}{\Gamma_2(\cos\alpha+\beta_2)}.
\end{equation}
We want to precisely estimate $F(\alpha, \theta)$ and its dependence
on $\sin ^2\alpha''$ in the range of $\alpha''$ from 0 to
$\alpha_j''$, where the subscript ``j'' always denotes the edge of
shell 1. Through equations (A-2) and (A-3), we can express $\theta$ in terms of
$\alpha$ only:
\begin{equation}
\theta=
\frac{\sin\alpha(1-\frac{\beta_2}{\beta_1}+\frac{\beta_2\delta
t}{R_1})}{\cos\alpha+\beta_2}.
\end{equation}
Then $F(\alpha, \theta)$ becomes $F(\alpha)$. Also, from equation (A-3),
express $\sin^2\alpha''$ in terms of $\alpha$ only:
\begin{equation}
\sin^2\alpha''=
\frac{\sin^2\alpha}{\sin^2\alpha+\Gamma_2^2(\cos\alpha+\beta_2)}.
\end{equation}
Therefore, we can plot $F(\alpha)$ numerically as a function of
$\sin^2\alpha''$.

The only thing left is to calculate the upper limit of $\alpha$. Equation
(A-2) gives
\begin{equation}
\tan\alpha_j'' =
\frac{\theta_j}{\Gamma_2(1-\frac{\beta_2}{\beta_1}+\frac{\beta_2\delta
t}{R_1})}.
\end{equation}
Denote $k=\tan\alpha_j ''$, and apply it onto equation (A-3) and square
both sides of (A-3). Then we get a quadratic equation of
$\cos\alpha_j$:
\begin{equation}
(1+k^2\Gamma_2^2)\cos^2\alpha_j+2\beta_2k^2\Gamma_2^2\cos\alpha_j+k^2\Gamma_2^2\beta_2^2-1=0,
\end{equation}
with roots
\begin{equation}
\cos\alpha_j=
\frac{-\beta_2k^2\Gamma_2^2 \pm \sqrt{k^2+1}}{1+k^2\Gamma_2^2}.
\end{equation}
The second root ($-$) can be ruled out, because when we put it back into
equation (A-3), the second root gives $k<0$, while equation (A-2) requires
$k>0$. Thus
\begin{equation}
\alpha_j=
\arccos\biggl(\frac{-\beta_2\Gamma_2^2k^2+\sqrt{k^2+1}}{1+k^2\Gamma_2^2}\biggr)
\end{equation}
is the sole root of the upper limit of $\alpha$.

We numerically plot $F(\alpha)$ vs. $\sin^2\alpha''$ for the
following model parameter space: $\frac{\Gamma_2}{\Gamma_1}$ ranges
from 0.05 to 10, $\frac{\delta t}{R_1/(2\Gamma_1^2)}$ ranges from 0 to 1000,
and $\theta_j$= 0.1. We find, for $\frac{\Gamma_2}{\Gamma_1} < 1$,
$F(\alpha)$ is approximately a linear, monotonically decreasing function of
$\sin^2\alpha''$. Its maximum is 1 and is at $\alpha=0$, and minimum is
always $>$ 0.01 and is at $\alpha=\alpha_j$. A smaller
$\frac{\Gamma_2}{\Gamma_1}$ or a larger $\delta t$
always gives a smaller $\alpha_j$, thus a minimum of
$F(\alpha)$ closer to 1.

This clearly shows that for $\frac{\Gamma_2}{\Gamma_1} < 1$, $F(\alpha)$ is very
weakly dependent on $\sin^2\alpha''$ thus can be taken out of the integral as a
constant $\sim$ 1.

For $\frac{\Gamma_2}{\Gamma_1}>1$, $F(\alpha)$ is not always on order of unity.
Since shell 2 is moving toward shell 1, the relativistic beaming effect is
important (Panaitescu 2007). When $\delta t$ is very small, shell 2 is
very close to shell 1, so that $\alpha_j''$ is larger than
$\frac{1}{\Gamma_{rel}}$, where $\Gamma_{rel} \simeq \frac{\Gamma_2}{2\Gamma_1}$
is the relative LF between two shells. In this case we find that at an angle
$\alpha'' < \alpha_j''$, $F(\alpha)$ starts to drop very sharply from on order
of unity to infinitely small, and that dropping-down angle is approximately
equal to $\frac{1}{\Gamma_{rel}}$. This means that, in the case of
$\frac{\Gamma_2}{\Gamma_1}>1$, $\sin^2{\alpha_j''}$ in
equation (14) should be replaced by a more accurate term
$\min(\sin^2{\alpha_j''}, \frac{1}{\Gamma_{rel}^2})$. However this is only
required when $\delta t$ is small so that $\frac{T_s-T_p}{R_1/(2\Gamma_1^2)} <100$.
Thus the more accurate flux ratio curves for $\frac{\Gamma_2}{\Gamma_1}>1$ in our
Figure 2 will be slightly flatter in the region of
$\frac{T_s-T_p}{R_1/(2\Gamma_1^2)} <100$ than the ones shown. But our
conclusion about a faster shell 2 is {\it not} affected.

\end{document}